\documentclass[aps,prd,preprint]{revtex4}
\usepackage{dcolumn}
\usepackage{amsmath}
\usepackage{color}

\usepackage{graphicx}
\usepackage{epstopdf}
\def\b{\beta}

\def\p{\pi}                     

\long \def \blockcomment #1\endcomment{}

\newcommand{\bee}{\begin{equation}}
\newcommand{\ee}{\end{equation}}
\newcommand{\beea}{\begin{eqnarray}}
\newcommand{\eea}{\end{eqnarray}}

\begin{document}
\title{Conformal or Walking? Monte Carlo renormalization group studies of  SU(3) gauge models with fundamental fermions}
\author{Anna Hasenfratz}
\email{anna@eotvos.colorado.edu}
\affiliation{Department of Physics,
University of Colorado, Boulder, CO 80309, USA}

\begin{abstract} Strongly coupled  gauge systems with many  fermions are important in many phenomenological models. 
 I use the 2-lattice matching  Monte Carlo renormalization group method  to study the fixed point structure and critical indexes 
of SU(3) gauge models with 8 and 12 flavors  of fundamental fermions.  With an improved renormalization group block transformation 
I am able to connect the perturbative and  confining regimes of the  $N_f=8$ flavor system, thus verifying its QCD-like nature. With $N_f=12$ flavors the data favor the existence 
of an infrared fixed point and conformal phase, though the results  are  also consistent with very slow walking.   I measure the anomalous mass dimension in both systems at several gauge couplings and find  that they are barely different from the free field value. 
\end{abstract}


\maketitle


\section{Introduction}

 The nature of the electroweak symmetry breaking is one of the most pressing questions of elementary particle physics today. Strongly coupled gauge-fermion systems, coupled to the electroweak Standard Model, could provide
  a dynamical mechanism for mass generation \cite{Hill:2002ap}.   Many of the interesting  models are asymptotically free theories based on an $SU(N_c)$ gauge group with different number of  fermions in various representation.  
 Systems with relatively few and/or low representation fermion species are typically QCD-like, exhibiting confinement and spontaneous chiral symmetry breaking ($\chi$SB). Models with many flavors or high representation fermions can develop an infrared fixed point (IRFP) in the gauge coupling \cite{Banks:1981nn}. These systems are conformal  and show neither confinement nor spontaneous chiral symmetry breaking. 
 Before the appearance of the IRFP, i.e. just below the conformal window, one might encounter  "walking" theories where the gauge coupling changes very slowly with the energy scale. 
 Walking is essential for technicolor models while conformal models are the basis of "unparticle" theories and might even be connected to string theories through the ADS/CFT conjecture. 
 
 Even though analytical semi-perturbative methods  give qualitative description of  the color and flavor dependence  of the  phase diagram and critical indexes \cite{Appelquist:1996dq,Appelquist:1998rb, Appelquist:1998xf,Sannino:2004qp,Dietrich:2006cm,Ryttov:2009yw,Braun:2009ns}, the strongly coupled systems should be studied  non-perturbatively. Several lattice groups have taken on this task recently
 \cite{Damgaard:1997ut,Heller:1997vh,Catterall:2007yx,Appelquist:2007hu,Shamir:2008pb,Deuzeman:2008sc,DelDebbio:2008zf,Catterall:2008qk,Fodor:2008hm,
DelDebbio:2008tv,DeGrand:2008kx,Jin:2008rc,Hietanen:2008mr,Appelquist:2009ty,Hietanen:2009az,Hasenfratz:2009ea,Hasenfratz:2009kz,Deuzeman:2009mh,Fodor:2009wk,DelDebbio:2009fd,Pica:2009hc,Bursa:2009we,
Sinclair:2009ec,Appelquist:2009ka,Bursa:2009tj,Fodor:2009rb,Fodor:2009ff,Jin:2009mc,DeGrand:2009hu,Kogut:2010cz,Yamada:2010wd}.
  These works investigate different gauge models, fermion numbers and representations, using different numerical methods. 
  The results, while exciting, are not always clear. The most interesting models frequently give contradictory signals and 
  require  more careful and detailed investigations than exist today. In this paper I consider the SU(3) gauge model 
  with various numbers of fundamental flavors and apply a Monte Carlo renormalization group (MCRG) technique to study the 
  running of the gauge coupling and the anomalous  dimension of the fermion mass. In  previous publications 
  \cite{Hasenfratz:2009ea,Hasenfratz:2009kz} I tested  the 2-lattice MCRG method on well understood systems and 
  carried out preliminary investigations for the $N_f=4$, 8, 12 and 16 flavor models. Here I improve on the original block transformation 
  and consider in detail  $N_f=8$ flavors, where practically every 
  calculation predicts QCD-like behavior, and   $N_f=12$ flavors, where the situation is much less clear \cite{Appelquist:2007hu,
Deuzeman:2008sc,Appelquist:2009ty,Deuzeman:2009mh,Fodor:2009rb,Fodor:2009ff,Jin:2009mc}. In the $N_f=8$ case I am 
able to connect the perturbative, weakly coupled regime to a confining system, thus verifying the expectations. I present results for the anomalous mass dimension 
at several gauge coupling values as well. 
My results for the $N_f=12$ case  favor the existence of an IRFP and conformal phase, though  I cannot exclude the possibility of very slow running.  
I find that the anomalous mass dimension is small, which is unexpected both at walking theories and at strongly coupled conformal systems.

Before discussing the details of the numerical calculation, in Sect.\ref{sect:Gen_description} I  briefly review the expected lattice phase diagram. 
Perhaps the cleanest way to distinguish QCD-like and conformal systems is to connect the perturbatively well understood weak coupling region to the strong coupling. This connection 
 can be explored through  the step scaling function, an integrated form 
 of the renormalization group (RG)  $\beta$ function \cite{Luscher:1992ny,Luscher:1992zx,Luscher:1993gh}. In Sect.\ref{sect:bare_step} I introduce the bare step scaling function \cite{Hasenfratz:1984bx} and discuss its properties in confining, conformal and walking-confining systems.
I use the 2-lattice matching Monte Carlo renormalization group  approach to calculate the (bare) step scaling function \cite{Hasenfratz:2009ea}.    
 In Sect.\ref{sect:method} I  summarize the method and introduce two new blocking schemes, both based on HYP smeared links \cite{Hasenfratz:2001hp}. These block transformations, referred to as HYP and HYP2,  integrate out the short distance ultraviolet fluctuations more effectively than the original transformation \cite{Swendsen:1981rb}  used in  Ref. \cite{Hasenfratz:1984bx,Bowler:1986rx} and work better in the strong coupling region.  Finally  Sect.\ref{sect:results} contains the numerical results for $N_f=$8 and 12 flavors.

\section{General description of the lattice phase diagram \label{sect:Gen_description}} 
  
\begin{figure}
\begin{center}
\vskip -1.8 cm
\includegraphics[width=0.8\textwidth,clip]{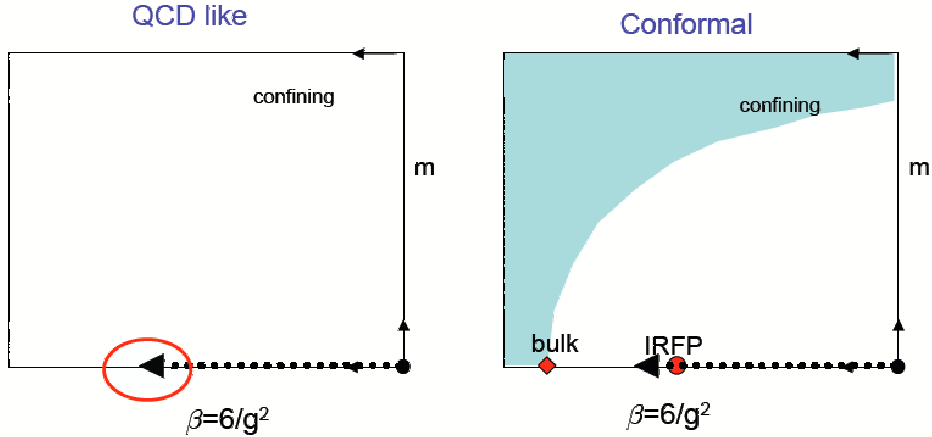}
\end{center}
\vskip -3.8 cm
\caption{The schematic phase diagram of QCD-like and conformal systems. }

\label{fig:phase_diagram}

\end{figure}

This section gives a brief  general  description of the lattice phase diagrams of QCD-like and conformal systems.
Some  of these  considerations have  been discussed in Ref.\cite{DeGrand:2009mt}.
Here, as frequently throughout this paper, I refer to any confining, chirally broken system as QCD-like, not distinguishing walking and QCD-like systems.
Figure \ref{fig:phase_diagram} shows  the schematic phase diagrams in  the fermion mass and gauge coupling phase space for  both  systems. At the perturbative fixed point at $g=0$, $m=0$  both operators are relevant in both cases, that is  $m$ and $g$ increase  as the energy scale is decreased. (Here, as in the rest of the paper, the flow is from the ultraviolet (UV) 
to the infrared (IR).)  At large $m$ the fermions decouple and one ends up with a pure gauge model, described
by an other perturbative fixed point at $g=0$, $m=\infty$. There  are no other physically relevant fixed points or phase boundaries in a QCD-like system, though spurious bulk transition sometimes show up in the strong coupling region. In a QCD-like system both the running gauge coupling and the running mass increase without bound as the energy scale decreases, unless the bare couplings are tuned to one of the perturbative FPs where continuum limits can be defined.

The conformal phase diagram is  more complex. In the chiral limit in addition to the $g=0$, $m=0$ ultraviolet fixed point (UVFP) there emerges a new fixed point, generally referred to as  Banks-Zaks or infrared fixed point. At the IRFP there is only one relevant operator, the fermion mass. The  gauge coupling is irrelevant and the running coupling approaches this IRFP, independent of the value of the bare coupling. 
 At large $m$ the fermions decouple, just as in the QCD-like case, so at heavy masses there is again a confining phase. This phase most likely extends all the way to the chiral limit at strong gauge coupling, as indicated by the shaded area of Figure \ref{fig:phase_diagram}b,  though it might end at finite $m$, $\beta=0$ \cite{Iwasaki:2003de}. There is no order parameter separating  the  phases at finite mass, and  likely there is no phase boundary either. 
 In the chiral limit the chiral condensate becomes an order parameter, so if the confining phase extends all the way to $m=0$, there has to be a phase boundary,  
  a bulk transition, in the gauge coupling. If the bulk transition is second order, it could play the role of the new ultraviolet fixed point (UVFP) suggested in Ref. \cite{Kaplan:2009kr}.   But independent of the existence and nature of the bulk transition,  there is no confinement or chiral symmetry breaking  around the IRFP.
 
 If a lattice simulation is to distinguish the two phase diagrams, it has to identify the main features of the two panels of Figure \ref{fig:phase_diagram}. Identifying  a confining phase at strong coupling is clearly not enough, neither is   
 identifying  a bulk phase transition  as lattice artifacts even in  QCD-like theories can lead to bulk transitions. The most promising approach appears to be to start in the weak coupling phase where connection to the perturbative regime can be made and connect it, via some renormalization group approach,  to the strong coupling region.  This process is indicated by the dashed arrows in Figure \ref{fig:phase_diagram}. If an IRFP is identified along the way, as is shown on the right panel, we can conclude that the theory has  a conformal phase. If no IRFP is found, one has to investigate further.  If large volume simulations can establish confinement and chiral symmetry breaking, we can conclude that the system is QCD-like. However if numerical simulations do not find confinement, one cannot make a definite conclusion. It is possible that the system is conformal  with an IRFP at some stronger coupling, or that it is QCD-like, but larger volumes are needed to firmly establish that.  Since at very strong couplings lattice artifacts can destroy any remnant of universality, it is possible that with a given lattice action and RG transformation the question cannot be resolved at all or at least not with the available computing resources, while an other lattice action and RG transformation might give a definitive result.

\section{The bare step scaling function \label{sect:bare_step}}

The renormalized running coupling and  the renormalization group $\beta$ function are scheme dependent, only  the  leading critical exponents at the fixed points are universal.  For the asymptotically free theories I consider here  the first two coefficients of the perturbative $\beta$ function are universal, the higher order terms depend both on the regularization and renormalization scheme. Similarly the existence of an infrared  fixed point in the gauge coupling of a conformal theory is universal, but its location within the critical $m=0$ surface is not. 
The anomalous mass dimension  is universal at an IRFP, but scheme and coupling dependent in a QCD-like system.

The step scaling function in Refs.\cite{Luscher:1992ny,Luscher:1992zx,Luscher:1993gh} is defined through  renormalized couplings, describing the change of the 
running coupling under a fixed change of  scale  in the continuum limit, that is around a UVFP. This step scaling function was designed to determine 
the renormalized coupling in QCD-like theories but it can be generalized to conformal systems as long as one is interested in the running of the coupling between the perturbative UVFP and the conformal IRFP. With the MCRG method I prefer to work with {\it bare}, rather than renormalized quantities, and introduce the differential bare step scaling function $s_b(g_0^2)$.   In this section I review the definition and some basic properties of the step scaling function both in QCD-like, conformal and walking theories. It is worthwhile to summarize some common properties of $s_b$:
\begin{enumerate}
\item $s_b(g_0 ^2)>0$ when the renormalization group $\beta(g_0 ^2)<0$ .
\item At a fixed point  $s_b(g_0 ^2)=0$.
\item In the vicinity of the perturbative  UVFP ($g_0=0$) at one loop level 
     \bee
  s_b(g_0 ^2;s=2) = -\,\frac{3 \,\rm{ln}(2)}{4\p^2} b_1\,\,\,\,\,\,
(\rm{1-loop}),
 \label{eq:db_pert}
  \ee
where $b_1<0$ is the first term of the RG $\beta$ function. 
\end{enumerate}
 The bare and
renormalized quantities can be connected, though I do not explore this possibility here. 

\subsection{The universality of the  bare step scaling function of a QCD-like lattice model}

The lattice formulation regularizes the continuum theory,  choosing a lattice action fixes the regularization scheme. The lattice model is formulated in terms of the bare couplings $\beta = 2 N_c /g_0^2$ and mass $m$. In the chiral limit there is no dimensional parameter in  a 
QCD-like system, but  dimensional transmutation generates the so called $\Lambda$ parameter
and the coupling dependence of every dimensional quantity, up to lattice artifacts, is determined   by  $\Lambda$. It is generally more convenient to work with dimensionless ratios, effectively setting the scale (or lattice spacing) by an arbitrarily chosen quantity, like the Sommer scale $ a = r_0^{\rm{phys}}/r_0^{\rm{lat}}$ \cite{Sommer:1993ce}. Then every dimension mass lattice quantity can be written as 
\bee
m^{\rm{lat}} = \frac{c}{r_0^{\rm{lat}}} +{\cal O}(a^2) \,,
\label{eq:scaling}
\ee
where the ${\cal O}(a^2)$ term represents scaling violations. As long as the lattice spacing is small and the scaling violations (lattice artifacts) are  controllable, the system is said to be in the scaling regime where cut-off independent physical predictions can be obtained.

The differential bare step scaling function of a system governed by a UVFP is defined as 
\bee
s_b(\beta;s)= \beta -\beta^{\prime}
\label{eq:sb}
\ee
 where the lattice spacing, defined through the arbitrary quantity that sets the scale, changes by a factor of $s$ between $\beta$ and $\beta^{\prime}$ 
 \bee
 a(\beta^{\prime}) = s a(\beta)\,.
 \ee 
 The scale change $s>1$ is arbitrary, but with the
 MCRG method I always consider $s=2$ and for simplicity in the following I drop the index $s$ in $s_b(\beta)$. 
 
    One can define a step scaling function using any  dimension mass lattice  quantity $m$ as
\beea
s_b^{(m)}(\beta)&=& \beta - \beta^{\prime\prime}\,, \cr
m^{\rm lat} (\beta^{\prime\prime}) &=& m^{\rm lat} (\beta)/s\,.
\eea
  According to Eq. \ref{eq:scaling}  
 \bee
 s_b^{(m)}(\beta) = s_b(\beta) +   {\cal O}(a)\, ,
 \ee
 i.e. the bare step scaling functions defined using different physical quantities are, up to lattice  artifacts,  identical in the scaling region of a UVFP. The same is true if we define $ s_b(\beta)$ using the scale dependence of the  running coupling in the Schrodinger functional formalism or by the MCRG prescription. As long as the lattice artifacts are small the different definitions of $s_b(\beta)$  should agree up to small corrections.

\begin{figure}
\begin{center}
\includegraphics[width=0.5\textwidth,clip]{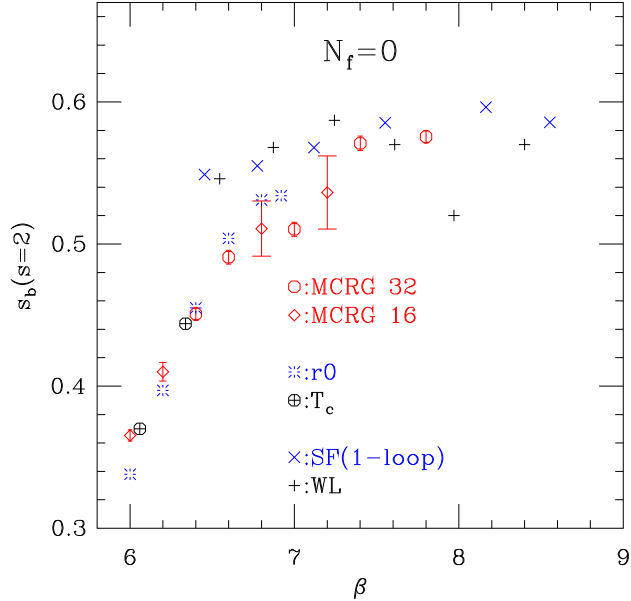}
\end{center}

\caption{The bare step scaling function $s_b(\b;s=2)$ for the pure gauge SU(3) system
as predicted by different methods. This figure is from Ref. \cite{Hasenfratz:2009ea}.
\label{fig:db_nf0}}

\end{figure}

Figure \ref{fig:db_nf0}, taken form Ref. \cite{Hasenfratz:2009ea},  shows the differential bare step scaling 
function for the $N_f=0$ pure gauge model, determined using the physical quantities $r_0$ \cite{Necco:2001xg} 
and the finite temperature phase transition $T_c$ \cite{Boyd:1996bx}, the Schroedinger functional method \cite{Luscher:1993gh} and the related Wilson loop method \cite{Bilgici:2009kh}, as well as the 2-lattice MCRG method on two different volumes. Errors are shown for the MCRG results only.  The agreement of the predictions in the weak coupling, as well as the agreement between $r_0$, $T_c$ and MCRG even at stronger couplings illustrate the universality of the step scaling function. The deviation of the Schroedinger functional and Wilson loop methods for $\beta<7.0$ is presumably due to lattice artifacts \cite{Necco:2001xg}. Since the Schroedinger functional data in Figure \ref{fig:db_nf0} are  only 1-loop improved, the relatively large difference is  not that surprising. 

\begin{figure}
\begin{center}
\includegraphics[width=0.5\textwidth,clip]{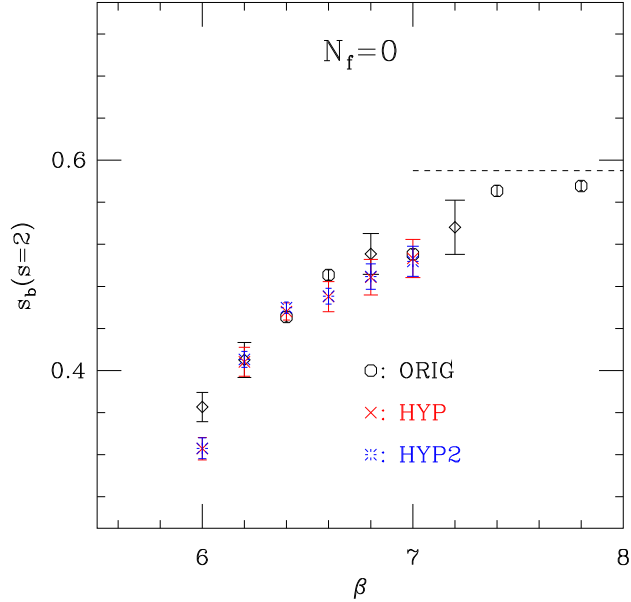}
\end{center}

\caption{The bare step scaling function $s_b(\b;s=2)$ for the pure gauge SU(3) system
as predicted by different RG transformations. The dashed line is the 1-loop perturbative prediction  $s_b^{(\rm{pert})}(s=2)=0.59$. 
\label{fig:db3_nf0}}

\end{figure}

In Figure \ref{fig:db3_nf0} I compare  the step scaling function as obtained with  3 different block transformations, the original one used in \cite{Hasenfratz:2009ea} and already shown in Figure \ref{fig:db_nf0}, and two new ones, based on one and two levels of HYP smearing. I will give the exact definition of these new RG transformations in Sect.\ref{sect:method}. The data in Figure \ref{fig:db3_nf0} are all from $16^4 \to 8^4$ volume matching and the agreement between the block transformations is almost perfect. That illustrates that all  3 RG transformations can be optimized, and  in the scaling regime they predict the same step scaling function with only small lattice artifacts.

The step scaling function deviates significantly form the 2-loop perturbative value in the range $\beta\in(5.7,6.5)$, though the agreement between the different RG methods and physical quantities imply that the system is still in the scaling region of the perturbative FP. This simply means that the lattice bare coupling is not a very good perturbative coupling. Replacing it, for example, with a tadpole improved coupling would reduce the difference between the perturbative and lattice measurements, effectively improving the convergence of the perturbative expansion. While tadpole improvement  would make the lattice result look more perturbative, in this case it is only a redefinition of the coupling. In order to keep the presentation as transparent as possible I rather use the original lattice couplings here.

\subsection{The bare step scaling function of a conformal lattice theory}

The situation is quite  different for models that develop an IRFP in the gauge coupling. First of all there is no dimensional transmutation in these systems. 
Correlation functions in the chiral limit show  power-like  decay (plus lattice corrections)
, not the exponential one of the QCD-like system; one cannot even talk about non-zero masses or finite  correlation length when $m=0$. The definition of $s_b(\beta)$ based on physical observables is not available now. The only option is to  define a step scaling function based on the running of a renormalized coupling or the renormalization group  flow. The definitions, however, do not lead to a universal quantity even in terms of the bare coupling. The location of the IRFP on the $m=0$ 
critical surface 
is scheme dependent, different RG transformations or  renormalized coupling definitions  can have different fixed points  and consequently different bare step scaling functions. 
\begin{figure}
\begin{center}
\includegraphics[width=0.5\textwidth,clip]{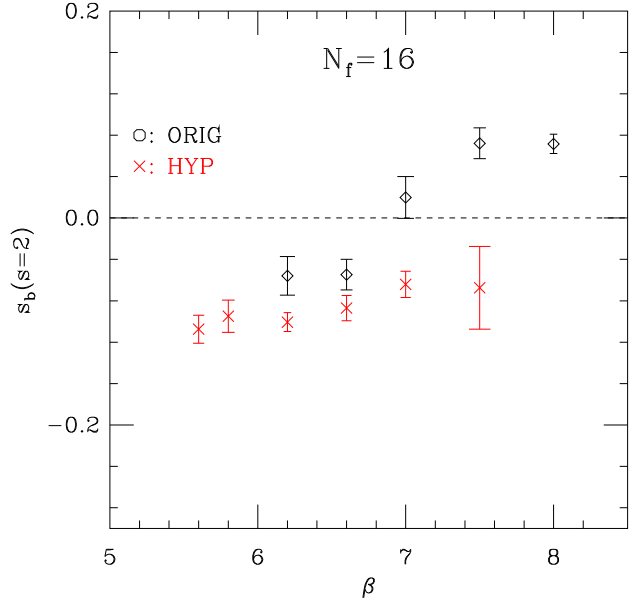}
\end{center}

\caption{The bare step scaling function $s_b(\b;s=2)$ for the 16 flavor system. The octagons and crosses correspond to the original and HYP RG transformations.}

\label{fig:db2_nf16}

\end{figure}

Figure \ref{fig:db2_nf16} illustrates this. It shows the step scaling function of the $N_f=16$ flavor model obtained with 2 different renormalization group transformations, the original and the HYP transformations.  In this theory the 2-loop $\beta$ function predicts an IRFP at fairly weak coupling, $6/g_0^2 \approx 12$. The data in Figure \ref{fig:db2_nf16} are all at stronger couplings. The HYP  RG transformation predicts a negative differential step scaling function, corresponding to positive renormalization group $\beta$ function in the investigated coupling range. The data are consistent with   
an IRFP at some weaker coupling. The original RG predicts an IRFP at  $6/g_0^2 \approx 7.0$.  As expected, the two RG transformations predict  different  step scaling functions and IRFPs. 

The renormalization group flows are  very different in  confining  and conformal systems. In the former the RG flows away from the UVFP with increasing rate, in the latter it flows into the IRFP with decreasing rate. In the chiral limit the 16 flavor system has no relevant operator in the vicinity of the IRFP. Most of the operators around the IRFP is expected to be strongly irrelevant. Perturbation theory suggests that the original gauge coupling, the operator that is relevant at the UVFP, has a small
scaling dimension at the IRFP, it is almost marginal.  The RG   the flow lines therefore should  run to a 1-dimensional line where only this near-marginal operator  is present, and slowly approach the IRFP along it. Between the UVFP and IRFP this 1-dimensional line is the RT of the UVFP, but there is no reason to expect that in the bare parameter space it does not continue beyond the IRFP, even if no other UVFP exists in the system. The negative $s_b$ step scaling function represents  the flow  along this extended RT. 

The anomalous dimension of the mass, related to the critical exponent in the still relevant mass direction, should  be universal at the IRFP. In Ref. \cite{Hasenfratz:2009ea} I found the anomalous mass dimension with the original blocking to be very small, 0.02(5). This suggests that the IRFP of the 16-flavor system is at weak coupling. Most likely the anomalous mass dimension is  equally small with the HYP blocking, though   I have not  done this calculation.

\subsection{The bare step scaling function of a walking lattice theory}

In a  walking model the RG $\beta$ function does not have an IRFP, the system is confining and chirally broken. Nevertheless the running of the coupling becomes
 very slow, the $\beta$
function develops a near zero. This situation is in-between the two scenarios discussed above: in the weak coupling region, where the running coupling is controlled by the perturbative UVFP,
one expects a universal step scaling function. When the $\beta$ function is near zero and the coupling runs very slowly, regularization effects can become significant, destroying the 
universal behavior.  A slowly running coupling of a walking theory can be difficult to distinguish from a conformal one where the coupling can   run slowly towards the IRFP, where it eventually stops. Lattice simulations encounter huge scale changes in a narrow coupling range, making it very difficult to distinguish the two scenarios.

\section{The  MCRG method \label{sect:method}}

The 2-lattice matching method measures the bare differential step scaling function by comparing expectation values measured on
blocked configurations. I have discussed the implementation
of the method in detail in Ref. \cite{Hasenfratz:2009ea}, here I just emphasize the main points and some new elements.

\subsection{The 2-lattice matching MCRG}

MCRG explores the phase and fixed point structure of lattice systems using a  real space RG block transformation of scale $s$. The block variables are 
 defined as some kind of  local average of
the original lattice variables. By integrating out the original
variables while keeping the block variables fixed, one removes the
ultraviolet fluctuations below the length scale $sa$. The lattice size and correlation length decrease by a factor $s$ at each blocking steps. 

The action describing the blocked system is usually complicated,
but the general properties of the  flow lines describing the blocked actions are 
quite universal.  Starting near
the critical surface the flow lines approach the fixed point in the irrelevant directions
but flow away in the relevant one(s). After a few RG steps the irrelevant
operators die out and the flow follows the  renormalized trajectory
(RT), independent of the original couplings. When the FP is governed by a single relevant operator, the RT is a 1-dimensional line. 
If we  identify
two  couplings, $\beta$ and $\beta^{\prime}$, that
lead to the same points along the 1-dimensional RT after repeated blocking,
 but one requires one less blocking
steps to do so,   the lattice correlation lengths at $\beta$ and $\beta^{\prime}$ must
differ by a factor of $s$.  According to Eq. \ref{eq:sb} that predicts the bare differential step scaling function $s_b(\beta)=\beta-\beta^{\prime}$. 

If the fixed point is governed by two relevant operators, like  the UVFP of asymptotically free theories, the RT is a two-dimensional surface and one has to tune   both
relevant operators to match the blocked actions. Tuning both operators can be avoided if one of the relevant operators is set to its fixed point value, for example $m=0$. In that case only the other relevant operator (the gauge coupling) flows and matching proceeds in one coupling only.

 It is not necessary to determine the blocked actions themselves to verify that they are identical.
It is sufficient to show that operator expectation values calculated with the two actions are identical. In the
matching procedure for the gauge coupling I generate configurations with the original action at couplings $\beta$ and $\beta^{\prime}$ in or very near the $m=0$ chiral limit, block
them as many times as possible and measure the expectation values of several operators on the blocked lattices. If all expectation values can be
matched by tuning the coupling $\beta^{\prime}$, I conclude that the blocked actions are identical. 

On the finite volume of a numerical simulation one can do only a few RG blocking steps and the method could break down if the 
RT is not reached within a step or two. Fortunately the location of the RT, just like the location of the FP, is not universal and different 
RG transformations have different RT trajectories. The "art" of MCRG matching is to find an RG transformation that
has a nearby RT so consistent matching can be made.

\subsection{ The ORIG, HYP and HYP2 block transformations}

The renormalized trajectory describes perfect actions, actions without lattice artifacts. While an RG transformation cannot get rid of long distance lattice artifacts, 
the more effective it is in integrating out the short distance UV fluctuations, the faster its flow approaches the RT. The original RG transformation introduced in 
\cite{Swendsen:1981rb,Hasenfratz:1984bx} and used in \cite{Bowler:1986rx,Hasenfratz:2009ea} is a good block transformation on smooth configurations but not particularly effective on coarser lattices. 
Here  I introduce two other RG transformations that work considerably better on coarse configurations.

  The blocked links of the original (ORIG) RG are defined as 
 \begin{equation}
V_{n,\mu}={\rm Proj \Big[}(1-\alpha_1)U_{n,\mu}U_{n+\mu,\mu}+\frac{\alpha_1}{6}\sum_{\nu\ne\mu}U_{n,\nu}U_{n+\nu,\mu}U_{n+\mu+\nu,\mu}U_{n+2\mu,\nu}^{\dagger} \Big]\,,\label{eq:block-trans}
\end{equation}
where ${\rm Proj}$ indicates projection to $SU(3)$ and the parameter $\alpha_1$ is used to optimize the block transformation at each  coupling separately .  
The  blocked links of the second RG transformation are built  from HYP smeared links $W^{\rm{HYP}}_{n,\mu}$  \cite{Hasenfratz:2001hp} as
\bee
V_{n,\mu}=W^{\rm HYP}_{n,\mu} W^{\rm HYP}_{n+\mu,\mu}\,.
\ee
HYP smearing has three free parameters. In this RG transformation I keep the inner two parameters as
they were set in the original HYP smearing, $\alpha_2=0.6$ and $\alpha_3=0.3$, and  use the last parameter, $\alpha_1$,
to optimize the blocking. I refer to this block transformation as HYP blocking. 
The third transformation is similar to the second except that the $W^{\rm{HYP}}_{n,\mu}$  links are twice HYP smeared.
As before, I fix the inner parameters, in this case to
$\alpha_2=\alpha_3=0.3$ and use the outer parameter to optimize the blocking. This is the  HYP2 block transformation.  

The choice of the inner parameters in the HYP and HYP2 blockings is rather arbitrary. In fact I tried several different values. It turns out that within reasonable limits the exact values of the inner parameters
are not very important, though the difference between the original, HYP and HYP2 transformations is quite significant.

\subsection{ The step scaling function with optimized  RG blocking \label{sect:univ_vs_opt}}

All three RG transformations I use have a free tunable parameter. I use this parameter to optimize the blocking at each gauge coupling individually.
For optimal blocking I require that the last 2 blocking levels predict the same step scaling function. That requirement can (almost) always be satisfied for any given operator, but
different operators might require different $\alpha_1$ parameter and predict different $s_b(\beta)$ scaling functions. The spread of the predictions (or the standard deviation of   $s_b(\beta)$
from the different operators) characterize the accuracy (goodness)  of the matching.

\begin{figure}
\begin{center}
\includegraphics[width=0.5\textwidth,clip]{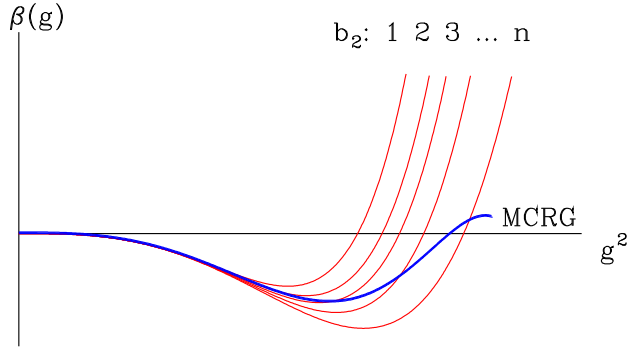}
\end{center}

\caption{The series of red lines indicate possible RG $\beta$ functions that differ in the non-universal coefficient of the 3-loop term, $b_2$. An optimized MCRG could pick a different RG 
transformation at each gauge coupling value, thus predicting the thick blue line as an "RG $\beta$ function" . }
\label{fig:betafn_vs_mcrg}

\end{figure}

An unintended consequence of this optimization is that the step scaling function is determined at each gauge coupling value with a different block transformation. 
As long as the step scaling 
function is universal, i.e. governed by a UVFP, this is not a problem, as is seen in Figure \ref{fig:db3_nf0}. On the other hand when universality is lost, like in the case of conformal or walking models, the step 
scaling function determined with optimized MCRG  does not correspond to any actual  fixed RG transformation. 
This is illustrated in Figure \ref{fig:betafn_vs_mcrg}.  
The thin (red) lines in the figure depict possible RG $\beta$ functions for the $N_f=16$ flavor model. They differ only in the coefficient $b_2$, the 3-loop term. 
The optimized MCRG can "pick" an RG with a different $b_2$ coefficient at each coupling value, 
predicting the thick blue line as the 
 $\beta$ function. While the MCRG prediction does not correspond to any specific RG transformation, its $\beta$ function cannot have a zero unless the RG transformations have one.
 In particular the step scaling functions in Figure  \ref{fig:db2_nf16} do not correspond to any particular RG transformation, but a negative value of $s_b(\beta)$  nevertheless indicate the existence of an IRFP.

One last note about the optimized RG transformations is in order before presenting the numerical results.  Not every block transformation has a fixed point. For example APE and 
HYP smearings disorder if $\alpha_1>0.75$ \cite{Bernard:1999kc}. The same is true for the original block transformation, as can be  seen from the analytical formulae in Ref.\cite{DeGrand:1995ji}, 
i.e. none of the block transformations I consider  have a FP if  $\alpha_1>0.75$.  
In practice the optimal RG transformation requires increased smearing, or larger  $\alpha_1$ parameter, on coarser lattices, and the matching eventually breaks down 
when the value of the optimal $\alpha_1$ approaches the stability limit. This will become clear in Sect.\ref{sect:results}.

\subsection{The anomalous mass dimension with the 2-lattice matching method \label{sect:anom_mass}}

The anomalous mass dimension at a fixed point can be calculated with the 2-lattice matching MCRG similarly to the step scaling function. At an IRFP the mass is the only  relevant operator, the gauge coupling is irrelevant. Matching expectation values at fixed, or even arbitrary, gauge couplings while varying the mass gives  pairs $(m_1,m_2)$ where the lattice correlation lengths differ by a factor of $s$,
\bee
\xi(m_1) = s \xi (m_2)\,\,.
\label{eq:mass_rel}
\ee
If $\xi(m) \sim m^{-1/y_m}$, Eq. \ref{eq:mass_rel} leads to 
\bee
m_2 = m_1 s^{y_m}\,\,,
\label{eq:m2_vs_m1}
\ee 
predicting the scaling dimension $y_m$ and the corresponding anomalous mass dimension 
\bee
\gamma_m= y_m-1\,\,.
\ee
 For matching in the mass the same set of operators can be used  as in the gauge coupling. I have illustrated this approach for the $N_f=16$ system in  Ref.\cite{Hasenfratz:2009ea}.  

In a system that has no  IRFP the gauge coupling remains relevant, matching  requires tuning both in the gauge coupling and mass, i.e. one needs to find pairs 
$(\beta,m_1;\beta^{\prime},m_2)$ that lead to matched blocked actions. While in principle it is possible to match in two couplings, in  practice it is easier to separate the two directions. In QCD-like or walking systems I first match the gauge couplings in the chiral limit, identifying a matched  pair $(\beta,\beta^{\prime})$ . This gives not only the shift in the gauge coupling under scale change $s$ but also predicts the optimal RG block transformation. Next I find matching pairs in the mass $(m_1,m_2)$ at the predicted  $(\beta,\beta^{\prime})$ values using the optimal block transformation. If matching is possible, the combination $(\beta,m_1;\beta^{\prime},m_2)$ corresponds to matched actions. Since the gauge coupling and mass operators do not mix, this 
2-step approach is valid at least for small masses. The breakdown of the matching is signaled when different operators predict different matching pairs. 

The general 2-step matching approach  can be used in conformal systems as well.  Since the gauge coupling can run slowly, it might not reach  the IRFP with the limited number of block transformations available. Matching in the mass then  is possible only if the gauge coupling is tuned as well.

\section{Simulation results \label{sect:results}}
The  numerical results of this paper were obtained with plaquette gauge action and nHYP smeared staggered fermions \cite{Hasenfratz:2007rf} with $16^4\to 8^4$ volume  matching.  
I have used 2-400 configurations separated by 10 molecular dynamics steps at each gauge coupling and matched 5 operators, the plaquette, the 3 6-link loops and a randomly chosen 8-link loop. 

\subsection{The $N_f=8$ flavor model}

The running coupling of the 8 flavor model has been studied with the Schrodinger functional method \cite{Appelquist:2007hu,Appelquist:2009ty}, and 
results consistent with perturbative scaling were found.
The finite temperature measurements in Ref. \cite{Deuzeman:2008sc}, and the p-regime  spectral measurements  and $\epsilon$-regime  Dirac spectrum   in Ref. \cite{Fodor:2009ff}   all found behavior  consistent with the perturbative UVFP as well. None of these works connected directly the perturbative 
 regime to a confining phase, a requirement I consider necessary to firmly distinguish the QCD-like and conformal phase diagrams of Figure \ref{fig:phase_diagram}.  
 Here I will close this gap and also present results for the anomalous mass dimension at several gauge coupling values.

\subsubsection{The step scaling function}

\begin{figure}
\begin{center}
\includegraphics[width=0.5\textwidth,clip]{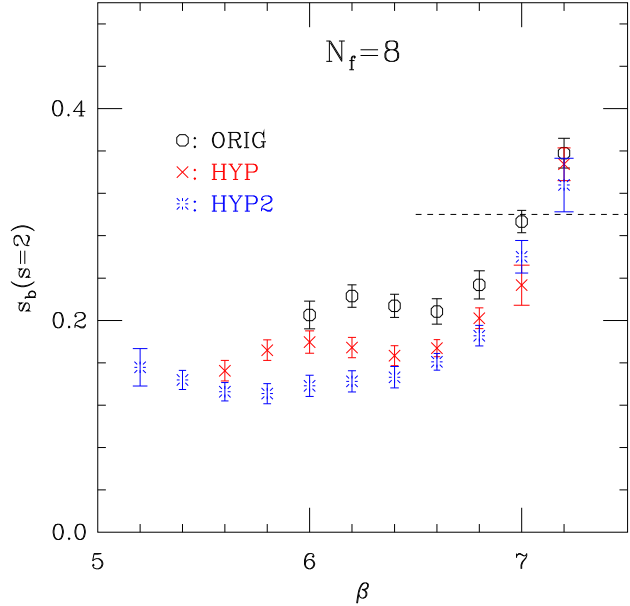}
\end{center}

\caption{The bare step scaling function $s_b(\b;s=2)$ for the 8 flavor system. The dashed line indicates the 1-loop perturbative prediction.}

\label{fig:db3_nf8}

\end{figure}

Figure \ref{fig:db3_nf8} shows the step scaling function as obtained by  the 3 RG transformations. All data  in the figure are from matching $16^4 \to 8^4$ volumes and the bare quark mass is $m=0.01$ matched to $m=0.02$. I have verified that at these small mass values the expectation values of the operators  used in the matching are, within statistical errors,  independent of the mass (see Ref.\cite{Hasenfratz:2009ea}), so these simulations can be considered to be in the chiral limit.
 Matching with the original transformation is possible only for $\beta\ge6.0$, while the HYP blocking breaks down at  $\beta \approx 5.6$, where   the optimal blocking parameter $\alpha_1$ exceeds the stability criterium $\alpha_1\le 0.75$.  The HYP2 blocking can be pushed considerably further, up to $\beta\ge 4.6$.  At the strongest gauge couplings  the configurations are very coarse, the plaquette at $\beta=4.6$ is 1.25 out of 3.0. Lattice artifacts are substantial and the matching breaks down  as different operators do not predict consistent values.  The difference between the 3 blocking schemes is 
  larger with 8 flavors than in the pure gauge case in Figure \ref{fig:db_nf0}, indicating larger lattice artifacts. Nevertheless in the range where  they are reliable, all three block transformations indicate $s_b(\beta)>0$, suggesting that the investigated gauge coupling range, $\beta \in (4.6,7.2)$,  is connected to the perturbative UVFP.  The next step is to establish confinement at least at the strongest coupling investigated. 
 
 \subsubsection{The static quark potential}

\begin{figure}
\begin{center}
\includegraphics[width=1.0\textwidth,clip]{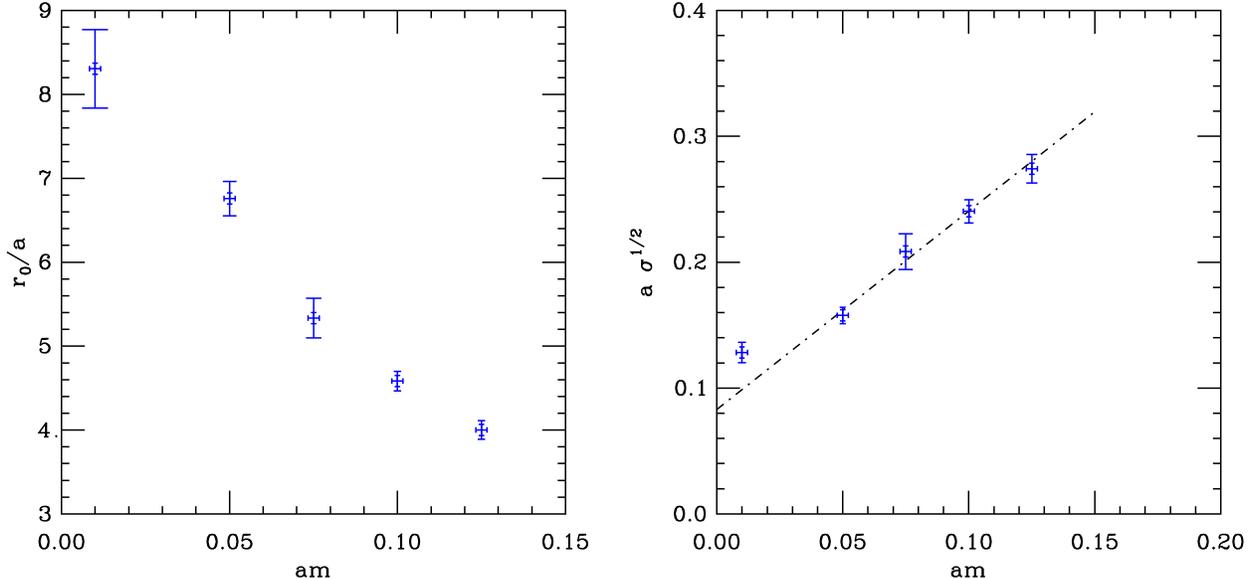}
\end{center}

\caption{$r_0/a$ and $a{\sqrt \sigma}$  as the function of the  quark mass at $\beta=4.8$ for the  $N_f=8$ flavor model }
\label{fig:potential_b48_nf8}

\end{figure}

 The largest volume I used in the MCRG matching is $16^4$. That is not large enough to verify confinement at $m=0.01$ and $\beta=4.8$ or 4.6. Instead of generating configurations on larger volumes, I decided to study the mass dependence of the static quark potential and the quantities $r_0$ and string tension $\sigma$ on the $16^4$ volumes. While the physical meaning of $r_0$ with $N_f=8$ flavors is probably very different from  2 or 2+1 flavor  QCD, it is still a good quantity to characterize the transition between the Coulomb and confining regions. 

As discussed in Sect.\ref{sect:Gen_description},  even conformal systems can show confinement at finite $m$. What distinguishes the conformal and QCD-like systems is what happens as the mass approaches  $m\to 0$. In the  chiral limit a conformal system should show only a Coulomb potential, while a QCD-like system is expected to have a non-vanishing  string tension as well. 
Figure \ref{fig:potential_b48_nf8} shows $r_0/a$ and $a{\sqrt  \sigma}$ at $\beta=4.8$ as the function of the quark mass, as obtained from the HYP smeared static potential on $16^4$ lattices \cite{Hasenfratz:2001tw} \footnote{In Ref. \cite{Hasenfratz:2001tw}  
we measured the static potential after HYP smearing because smearing greatly reduced the statistical fluctuations at long distances. In light of the MCRG discussion of the present paper
this is equivalent  to measuring the static potential on  an RG  blocked lattice.}.  The values of $r_0/a$ on the left panel indicate that finite volume effects are under control for $m\ge0.075$ where $L/r_0>3.0$. $a{\sqrt \sigma} $ on the right panel  
shows a linear dependence on the mass for $m\ge0.075$ and extrapolates to a finite value in the chiral limit. Data at $\beta=4.6$ are similar, predict a non-vanishing string tension at $m=0$. The 8-flavor model is confining at these gauge couplings, the renormalization group calculation of the step scaling function indeed connects the perturbative weak coupling region to the confining regime.

\subsubsection{The anomalous mass dimension \label{sect:nf8_anom_mass}}

\begin{figure}
\begin{center}
\includegraphics[width=0.5\textwidth,clip]{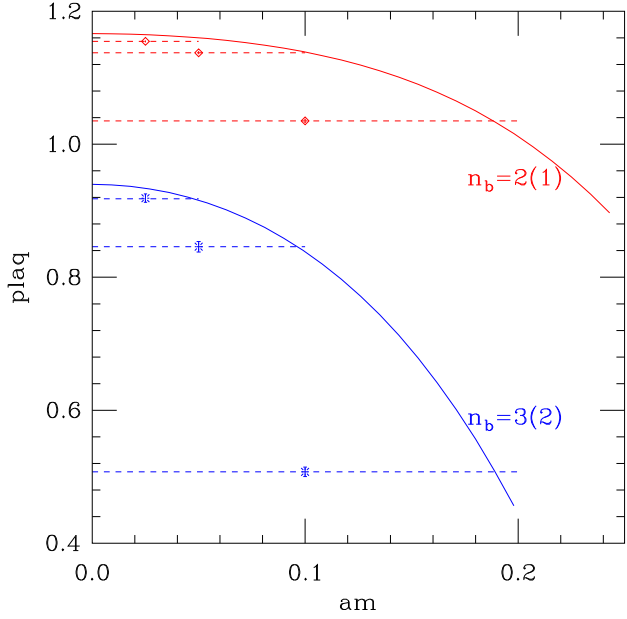}
\end{center}

\caption{Matching of the plaquette  in the mass at $(\beta=5.0,\beta^{\prime}=4.81)$. The upper red and lower blue  curves show the plaquette after $n_b=1$ and 2  blocking steps on the $8^4$, $\beta=4.81$ lattices. These are matched to the $n_b=2$ and 3 times blocked $16^4$, $\beta=5.0$ plaquette values (red diamonds and blue bursts). The horizontal lines indicate the matching.  }

\label{fig:mass_matching_nf8_b5}

\end{figure}

\begin{figure}
\begin{center}
\includegraphics[width=0.5\textwidth,clip]{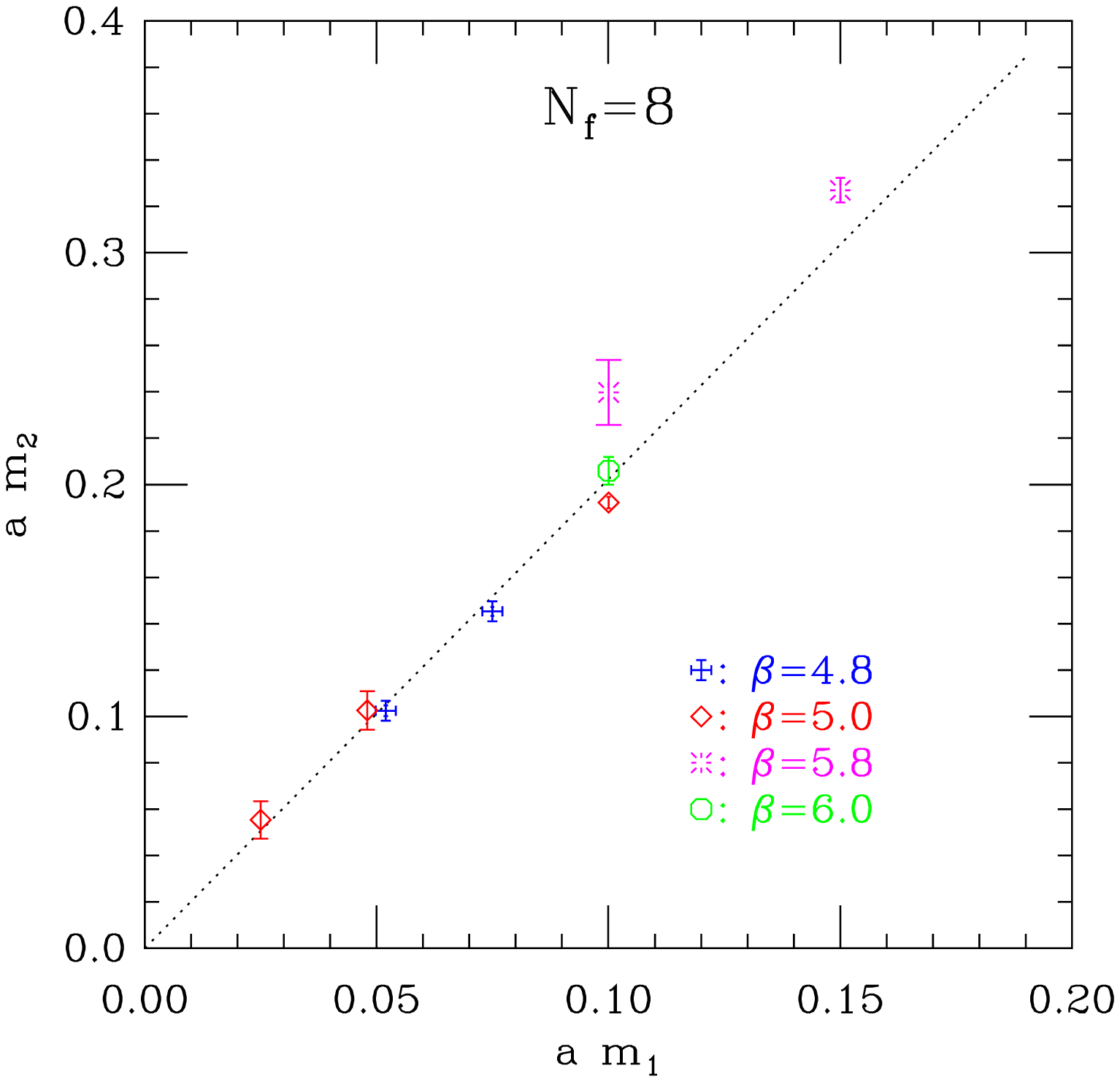}
\end{center}

\caption{Matching masses for the 8 flavor system.  In principle the slope of the matched mass pairs could be different at each gauge coupling, the result indicates that is not the case here. The data predicts, within error, a vanishing anomalous dimension.}

\label{fig:m2_vs_m1_nf8}

\end{figure}

An important quantity in conformal and walking theories is the anomalous dimension of the  mass. In systems where the gauge coupling is relevant, the 2-lattice matching MCRG requires
tuning both the gauge coupling and quark mass.  To simplify the numerical task  first I tune the gauge coupling in the chiral limit, and use these values in the matching of the mass. 
This procedure is correct for small masses. If/when it  breaks down for larger masses,  different operators predict different matching values, clearly signaling the breakdown.
The matching is illustrated in Figure \ref{fig:mass_matching_nf8_b5} for  $\beta=5.0$. In the chiral limit I found that the matching gauge coupling is $\beta^{\prime}=4.81(1)$ and the optimal blocking parameter is 
$\alpha_1=0.50(1)$.  Figure \ref{fig:mass_matching_nf8_b5} shows the plaquette values on $8^4$ lattices at $\beta=4.81$ after $n_b=1$ (upper, red curve) and $n_b=2$ (lower, blue curve) blocking steps as the function of 
the mass.  
This is matched with the $16^4$ simulations at $\beta=5.0$, $m=0.025$,  0.05 and 0.10. The red diamonds and blue bursts show the corresponding plaquette values after $n_b=2$ and 3 
blocking steps, and the  horizontal lines indicate the matching between the $16^4$ and $8^4$ data. The RG blocking parameter is  $\alpha_1=0.50(1)$,  the optimal one found in the chiral limit.  The 2 blocking levels predict consistent matching values, showing that the optimization at $m=0$
is valid at finite mass as well.  The other four operators show similar results, predicting consistent matched $(m_1,m_2)$ pairs. As 
Figure \ref{fig:mass_matching_nf8_b5}  shows the mass dependence of my observables is weak at small quark masses, $m_1=0.025$ is about the smallest value where the matching with gauge operators   works.

I have repeated the mass matching at several other gauge coupling pairs: $(\beta,\beta^{\prime})=(4.8,4.55)$, (5.8,5.67), and (6.0,5.81). Figure \ref{fig:m2_vs_m1_nf8} shows the matching mass pairs for all
four sets.  The anomalous dimension in a confining system  depends on the gauge coupling, and in principle the four gauge coupling pairs could predict different anomalous dimensions. This appears not to be the case, in Figure \ref{fig:m2_vs_m1_nf8} all data points are consistent with the same linear fit predicting $y_m=0.99(2)$ or anomalous dimension
\bee
\gamma_m=0.00(2)\,\,,\,\,\beta=4.8-6.0 \,.
\ee

In a chirally broken system the anomalous dimension is expected to be large, $\gamma_m= {\cal O}(1)$. The consistently small value I found here  indicates  that the simulations have not reached the scale of 
chiral symmetry breaking  yet.

\subsection{The $N_f=12$ flavor model}

My calculation of the $N_f=12$ flavor system mirrors the 8-flavor case. I use nHYP smeared staggered fermions and match $16^4 \to 8^4$ volumes with the three RG block transformations I already introduced.

\subsubsection{The step scaling function}

\begin{figure}
\begin{center}
\includegraphics[width=0.5\textwidth,clip]{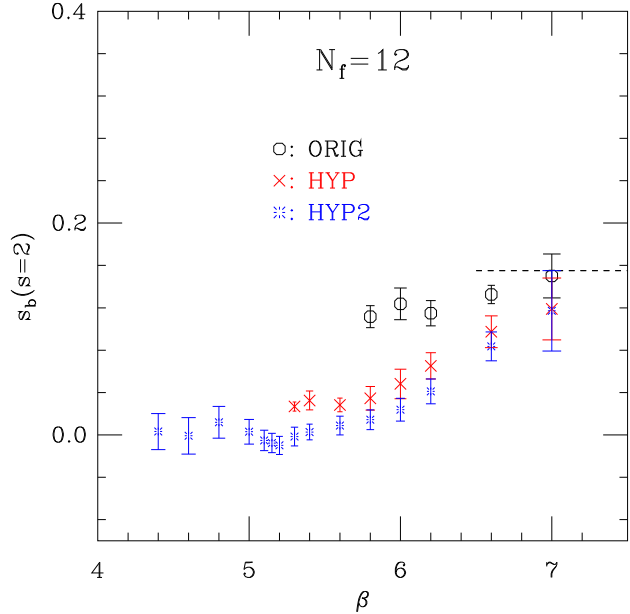}
\end{center}

\caption{The bare step scaling function $s_b(\b;s=2)$ for the 12 flavor system.}

\label{fig:db3_nf12}

\end{figure}

Figure \ref{fig:db3_nf12} shows the differential bare step scaling function of the $N_f=12$ flavor model. Just like in the 8-flavor case the original blocking breaks down fairly early, the HYP blocking works for $\beta\ge5.2$, while the HYP2 blocking can be pushed all the way to $\beta=4.4$.   The predicted step scaling functions both with HYP and HYP2 blockings differ significantly from the 1-loop perturbative value $s_b^{\rm 1-loop} = 0.16$. The HYP2 transformation predicts $s_b(\beta)=0$ within statistical errors in the range $\beta\in(4.4,5.6)$. As discussed in Sect.\ref{sect:univ_vs_opt} that does not necessarily  mean an RG $\beta$ function that vanishes across this coupling range, rather that  there exist  RG transformations  that have a zero  or near zero  in this range and the optimization procedure picks the  RG with a  (near) zero at or very near the investigated gauge coupling. 
 My data for now cannot distinguish the two possibilities: a true zero of the RG $\beta$ function, i.e. an IRFP, or a near zero of a walking theory.
  It is possible that at stronger gauge couplings the optimal RG transformation has a negative $s_b<0$ and therefore an IRFP, but with the present action I cannot test that.  The configurations are already too coarse and    the simulations get increasingly expensive at strong coupling, while universality becomes questionable as well.  Repeating the 2-lattice matching at the present gauge coupling range but on larger volumes could help. On $32^4$ lattices  one more blocking step  is possible. This will reduce the systematical errors considerably and one might even be able to study a limited but finite coupling range with the same RG transformation. This study is under way.

 \subsubsection{The static quark potential}
 
 The mass dependence of the static quark potential could distinguish conformal and confining systems. If the string tension measured at finite quark mass  extrapolates to a finite value in the chiral limit, the system is confining. If it extrapolates to zero, the system is  conformal. I have tried this analysis at 
 $\beta=4.2$, 4.4 and 4.6, but my results are inconclusive. The scale changes very rapidly with the coupling, but $16^4$ volumes are too small even at $\beta=4.2$ to reliably measure the string tension unless the mass is above $am>0.10$. At that point extrapolation to the chiral limit becomes 
 questionable. The configurations are too coarse to trust results at much stronger couplings and
  I did not pursue the static potential any further. Again, simulations on larger volumes would be useful.

\subsubsection{The anomalous mass dimension}

\begin{figure}
\begin{center}
\includegraphics[width=0.5\textwidth,clip]{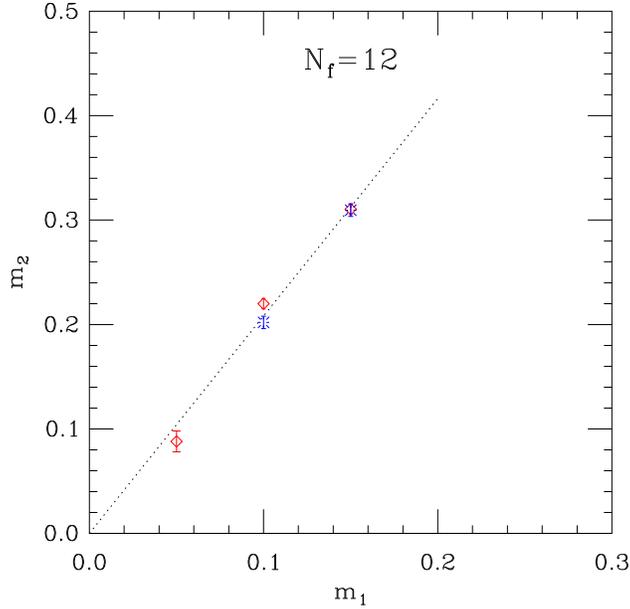}
\end{center}

\caption{Matching masses for the 12 flavor system. Red diamonds: $\beta=5.0$;blue bursts: $\beta=5.1$.}

\label{fig:m2_vs_m1_nf12}

\end{figure}

Strongly coupled conformal systems are expected to have large anomalous dimensions. Results for the step scaling function suggests the existence of an IRFP, and  the HYP2 transformation  predicts $\beta=\beta^{\prime}$ for $\beta \le 5.6$. I calculated matching mass pairs $(m_1,m_2)$ at $\beta=5.0$ and 5.1 following the procedure outlined in Sects.\ref{sect:anom_mass} and \ref{sect:nf8_anom_mass}. The results, shown in Figure \ref{fig:m2_vs_m1_nf12},  again indicate a very small anomalous dimension: $y_m=1.06(2)$. Other likely conformal systems also show small, though somewhat larger 
anomalous dimensions \cite{Bursa:2009we,DeGrand:2009hu}. I can imagine two possible explanations for my result. First, that the scaling regime in the mass is very small and the $m=0.05-0.15$ mass range is already outside of the scaling window. The other possibility is that the $N_f=12$ system  at $\beta=5.0-5.1$  is not strongly coupled. If it is walking, the chiral symmetry breaking scale is not reached at these gauge couplings. If it is conformal, its IRFP is not strongly coupled.  It would be very interesting to test models with $N_f=9-15$ flavors and check if those show an IRFP or walking with possibly larger anomalous dimensions. 

\section{Summary}

I have studied the renormalization group properties of the  $N_f=8$ and 12 flavor SU(3) gauge models with the 2-lattice matching technique.  Using improved block transformations I have been able to 
study these systems deeper in the strong coupling limit than in previous publications. For both models I calculated the bare step scaling function in the chiral limit and the anomalous mass dimension at several gauge coupling values. My results for the  8 flavor case are consistent with other methods, implying a confinining, chirally broken system, though the anomalous dimensions at the gauge couplings I considered are still small, free-field like. The results for the 12 flavor case favor the existence of an infrared fixed point, though it is also consistent with very slow walking. The anomalous mass dimensions are again small, a somewhat puzzling result. 

The numerical results presented in this paper are based on $16^4 \to 8^4$ volume matching. Repeating the calculation on larger, $32^4\to 16^4$ volumes  would considerably reduce the systematical errors and clarify many of the still open questions in these models.

\section{Acknowledgment }

I thank  Prof. J. Kuti for extensive discussions about the many intricate properties of conformal and walking models and the techniques used to study them. I thank the participants of the workshops  "Universe in a Box", Lorentz Center, Leiden, NL, August 2009,  "Strongly Coupled System", Boston, November 2009 and "New Applications of the Renormalization Group Method in Nuclear,
Particle and Condensed Matter Physics", INT, Seattle, February 2010,  for discussions, comments  and invaluable questions about the MCRG matching method. The numerical calculations of this work were carried out on the HEP-TH computer cluster at the University of Colorado and at the USQCD Kaon cluster at FNAL. 
This research was partially supported by the US Department of Energy. 

\bibliographystyle{apsrev}
\bibliography{lattice}

\end{document}